\documentclass[pre,aps,twocolumn]{revtex4-1}
\usepackage{graphicx}
\usepackage{amsmath}
\usepackage{amssymb}
\usepackage{color}
\usepackage{natbib}
\usepackage{epstopdf}
\def\strutdepth{\dp\strutbox}
\def\nw#1{\strut\vadjust{\kern-\strutdepth\vtop to0pt{\vss\hbox to\hsize
{\hskip\hsize\hskip5pt$\leftarrow$\hss\strut}}}{\em #1}}

\def\Re{{\rm Re}}
\def\Bo{{\rm Bo}}
\def\Oh{{\rm Oh}}

\def\Ca{{\rm Ca}}

\begin{document}

\title{Oscillating and star-shaped drops levitated by an airflow}

\author{Wilco Bouwhuis$^{1}$, Koen G. Winkels$^{1}$, Ivo R. Peters$^{1,2}$,\\ Philippe Brunet$^{3}$, Devaraj van der Meer$^{1}$, and Jacco H. Snoeijer$^{1}$}
\affiliation{$^{1}$Physics of Fluids Group, Faculty of Science and Technology, University of Twente, 7500 AE Enschede, The Netherlands \\
$^{2}$James Franck Institute, University of Chicago, Chicago, Illinois 60637, USA \\
$^{3}$Laboratoire Mati\`ere et Syst\`emes Complexes UMR CNRS 7057, 10 rue Alice Domon et L\'eonie Duquet 75205 Paris Cedex 13, France}

\date{\today}

\begin{abstract}

We investigate the spontaneous oscillations of drops levitated above an air cushion, eventually inducing a breaking of axisymmetry and the appearance of `star drops'. This is strongly reminiscent of the Leidenfrost stars that are observed for drops floating above a hot substrate. The key advantage of this work is that we inject the airflow at a constant rate below the drop, thus eliminating thermal effects and allowing for a better control of the flow rate. We perform experiments with drops of different viscosities and observe stable states, oscillations and chimney instabilities. We find that for a given drop size the instability appears above a critical flow rate, where the latter is largest for small drops. All these observations are reproduced by numerical simulations, where we treat the drop using potential flow and the gas as a viscous lubrication layer. Qualitatively, the onset of instability agrees with the experimental results, although the typical flow rates are too large by a factor 10. Our results demonstrate that thermal effects are not important for the formation of star drops, and strongly suggest a purely hydrodynamic mechanism for the formation of Leidenfrost stars.

\end{abstract}

\maketitle

\section{Introduction}

\begin{figure}[htp!]
\centering
\includegraphics[width=8.0cm]{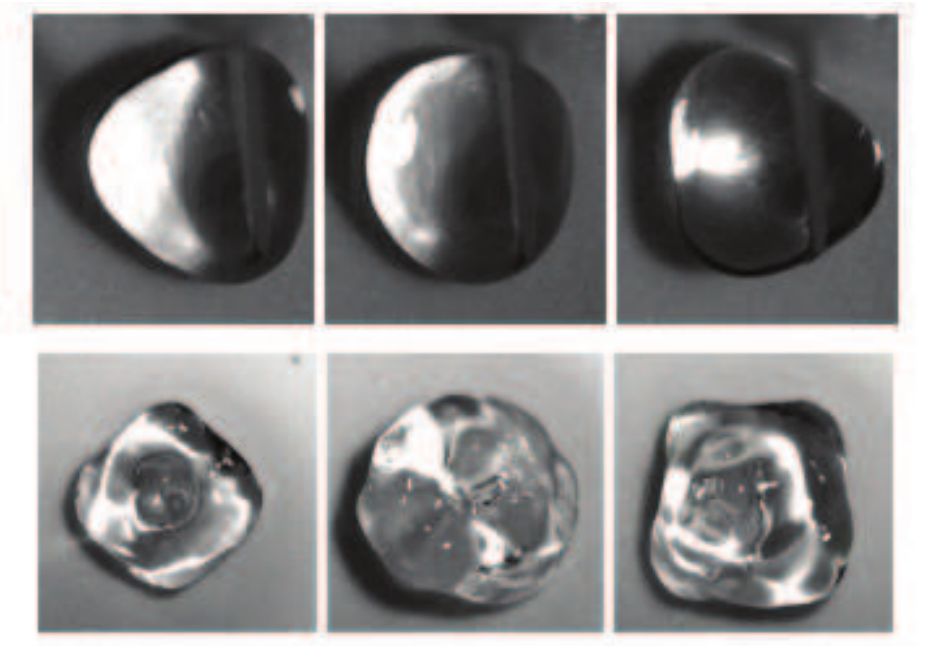}
\caption{Star drops levitated by a steady (i.e. non-pulsating) airflow. Top: mode $n$=3; bottom: mode $n$=4. Figure from Ref. \cite{Brunet11}.}
\label{nonpulsedaircushionmodes}
\end{figure}

Drops of water can levitate above a very hot plate due to the so-called `Leidenfrost' effect \cite{Leidenfrost:1756,Quere13}. In this situation, drops float on a thin layer of water vapor that results from evaporation in between the hot substrate and the drop. The shape and dynamics of the vapor layer can be quite complex \cite{Nagel12} and can be used to move liquid along a surface with the help of unevenly textured substrates \cite{Linke06,Lagubeau11,Wurger11}. 
Under some conditions, drops spontaneously start to oscillate and develop `star-shapes' or `faceted shapes' \cite{Japonais84,Japonais85,Holter52,Strieretal00,Aranson08}. Recently, it has been found that this phenomenon does not only occur in the case of Leidenfrost drops, but also for drops levitating on a steady and ascending uniform airflow at room temperature \cite{Brunet11}. Fig. \ref{nonpulsedaircushionmodes} shows examples of levitating star-drops obtained with water, taken from Ref. \cite{Brunet11}. The origin of the oscillatory instability has remained unclear, but the striking similarities with the Leidenfrost stars suggest a common mechanism for both, based only on hydrodynamics and free-surface dynamics, without invoking any thermal effects. 

Drops with faceted shapes have been observed in various systems with a periodic forcing of frequency close to the eigenmodes of the drop. Such drop shapes arise for drops on vertically vibrated hydrophobic substrates \cite{Noblin05,Okada06}, acoustically levitated drops with low-frequency modulated pressure \cite{acoustic_levit10}, liquid metal drops subjected to an oscillating magnetic field \cite{Fautrelle05}, or drops on a pulsating air cushion \cite{Papoular97,Perez99}. Using simple arguments \cite{Japonais96}, the appearance of these stars can be explained by the temporal modulation of the eigenfrequency of the drop, due to the external forcing, thus inducing a parametric instability. This suggests the following scenario for the formation of stars in a \emph{steady} ascending airflow: A first instability leads to a vertical oscillation of the drop, which through a secondary, parametric instability leads to the formation of (period doubled) oscillating stars.

Rayleigh and Lamb \cite{Rayleigh:1879} already predicted that for small enough deformations and for inviscid spherical drops, the resonance frequencies of the drops are given by:
\begin{equation}
f_n  = \frac{1}{2 \pi} \left( \frac{n (n-1) (n+2) \gamma}{\rho_l R^3} \right)^{1/2},
\label{eq:ray}
\end{equation}
\noindent where $f_n$ stands for the resonance frequency of the $n^{th}$ mode of oscillation, $R$ is the radius, $\gamma$ and $\rho_l$ are the liquid surface tension and density, respectively.
When the drop shape is different from the ideal spherical case, the resonance frequencies are modified with much more complex expressions, but in the case of a liquid puddle of radius $R$ much larger than the averaged drop height $H_d$, the eigenfrequencies take the following simple expression~\cite{Japonais96} :
\begin{equation}
f_n=\frac{1}{2\pi}\left(\frac{n(n^2-1)\gamma}{\rho_l R^3}\right)^{\frac{1}{2}},
\label{eq:cylindricalRayleigh}
\end{equation}
\noindent where $n$ is now the number of lobes on the drop in the azimuthal direction. Note that in practice, the frequencies predicted by eq. (\ref{eq:ray}) and (\ref{eq:cylindricalRayleigh}) are very similar. Thus it becomes clear that a parametric instability should occur when the drop radius is modulated in time. The same happens when due to a periodic external forcing, the drop stands in a time-periodic acceleration field. In that case the height of the cylindrical liquid puddle $H_d$ also varies periodically, and for a non-wetting condition (contact-angle close to 180$^{\circ}$) this height is simply equal to twice the effective capillary length $\ell_c = \sqrt{\gamma/(\rho_l a)}$, $a$ being the instantaneous acceleration (without forcing, $a$ is equal to the gravity constant $g$). By volume conservation, a time dependence of $H_d$ results into an oscillation of the radius $R$. Assuming small deformations, $R$ will have the same time-periodicity as the external forcing. Then, star-shaped oscillations by parametric forcing typically display a frequency equal to half of the driving (vertical oscillation) frequency \cite{Japonais96}.

In the case of a steady, non-pulsating air cushion or Leidenfrost levitation, the key question is to identify the origin of the vertical oscillations: What is the mechanism that induces a time-periodic instability, which in turn gives rise to vertical oscillations of the drop center-of-mass and shape? Once the origin of this instability is explained, the appearance of star drops is likely to originate from the parametric instability as stated above.
Recent experiments of star drops levitated on a continuous flow air cushion (Fig.~\ref{nonpulsedaircushionmodes}) suggest that these star-drops do not result from a temperature gradient-induced instability, contrary to what was previously hypothesized  \cite{Japonais94}. Apart from the oscillatory instability, a levitated drop can develop a `chimney', for which an air bubble develops below the drop and pierces through the center of the drop~\cite{Biance03}. This phenomenon has been explained theoretically from a breakdown of steady solutions~\cite{Duchemin05,SnoeijerPRE:2009}. Interestingly, the numerics for very viscous drops did not display any oscillatory instability. Therefore, the determination of the mechanisms for oscillations requires a more complex numerical scheme than those of Refs. \cite{Duchemin05,SnoeijerPRE:2009}.

\begin{figure}[htp!]
\centering
\includegraphics[width=5.0cm, height=5.0cm]{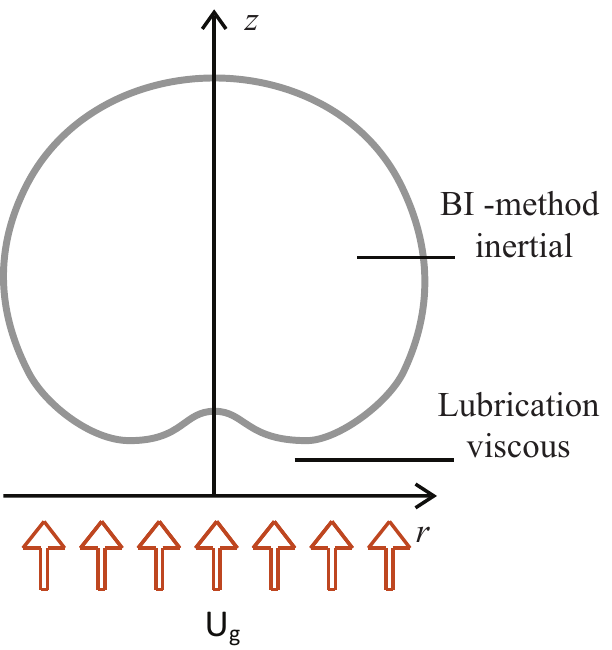}
\caption{Numerical implementation of the drop levitated by an airflow with uniform upward flow velocity $U_g$. The numerics consist of a coupling between the Boundary Integral method for the inviscid drop, and the lubrication approximation for the airflow beneath the drop. The flow inside the drop is assumed to be a potential flow; the flow at the bottom of the drop is a viscous flow, in which inertial effects are neglected.}
\label{fig:method}
\end{figure}

In this paper, we experimentally and numerically study drops levitated by an air-cushion, focusing on the instability to chimney formation, oscillations and star drops. The experiments consist of a significantly improved variant of that in Ref. \cite{Brunet11}, where we now can determine the threshold of instabilities with good accuracy. For the numerics, the proximity of the cushion to the drop calls for a method capable of accurately describing the gas-liquid interface, which leads us to employing an inviscid Boundary Integral method for the description of the drop. Inspired by the success of lubrication models in providing steady solutions for the drop shape we use a lubrication approximation for the airflow below the drop (Fig. \ref{fig:method}). This coupling has also been applied to simulate the impact of liquid drops on solid plates, and appeared to be successful in the regimes of both small and large impact velocities \cite{Bouwhuis}. The numerical implementation of the drop is completely axisymmetric and aims to explain the appearance of up-down oscillations for the drop.

The paper is organized as follows: In section \ref{sec:Experimental setup}, we present the setup we used to obtain the oscillating levitated drops experimentally, for liquids of different viscosities. Results of these experiments are shown in section \ref{sec:Experimental results}. Then, we describe the numerical scheme in detail (section \ref{sec:Numerical method}), and show the different regimes exhibited by the model (section \ref{sec:Numerical results}). In the last section, we conclude on these results.

\section{Experimental setup} \label{sec:Experimental setup}

\begin{figure}[htp!]
	\centering
	\includegraphics[width=0.4 \textwidth]{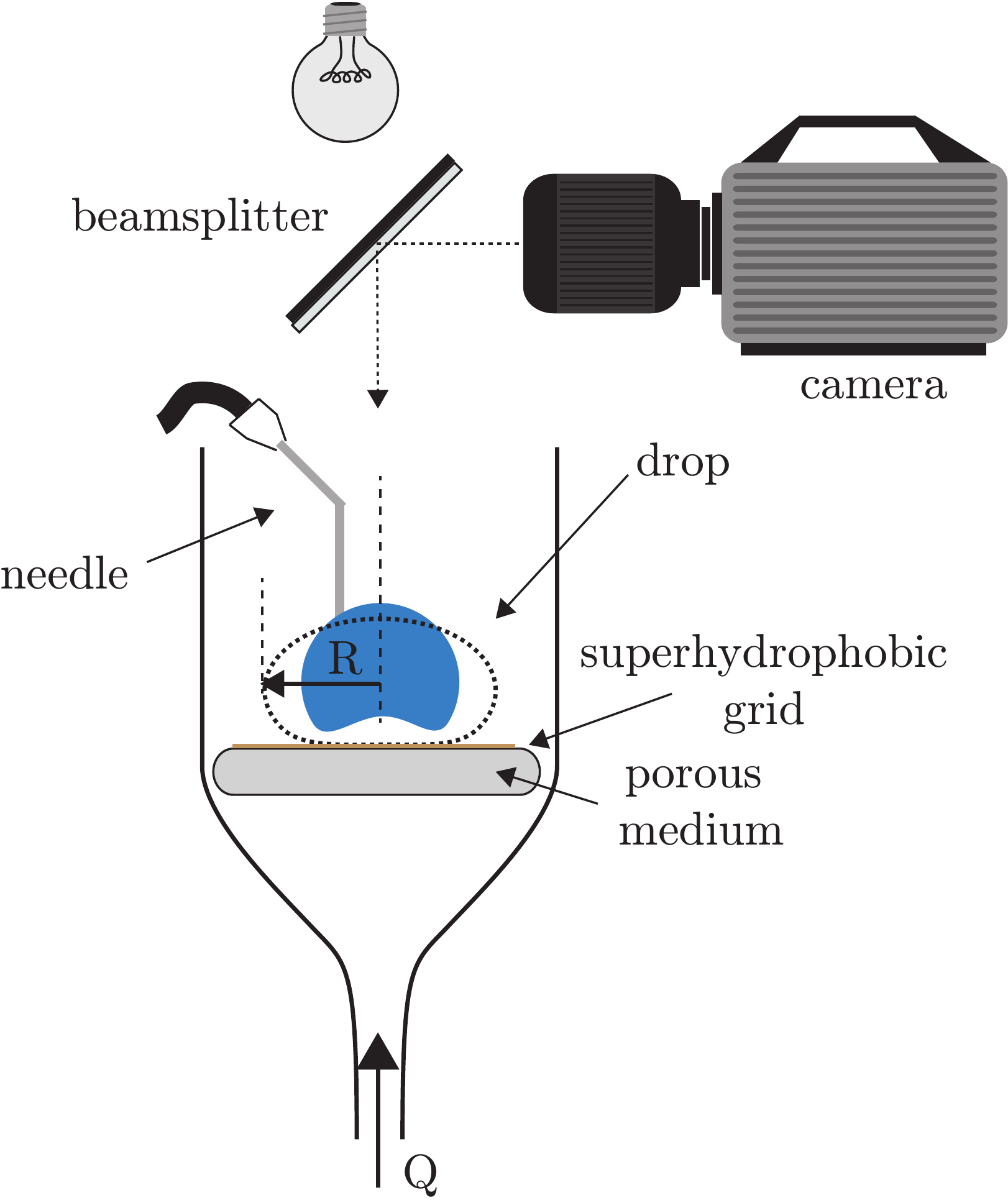}
  \caption{Sketch of the experimental setup. Illumination and camera view are obtained using a beamsplitter. A flow rate $Q$ is prescribed through a porous medium. Since the levitated drop is very mobile, it is held in position by a needle, which also supplies the liquid.}
   \label{fig:Problemsketch}
\end{figure}

It is well known that in case of Leidenfrost drops, the drops are levitated by a vapor layer. The vapor, coming directly from the drop, generates a cushioning layer for levitation due to the build up of a lubrication pressure between the lower part of the drop and the substrate. To avoid temperature effects and to directly control the gas flux in the layer, another experimental method was introduced in Ref.~\cite{Brunet11}. In this experimental method the air cushion is created by an ascending airflow (Fig.~\ref{fig:Problemsketch}). The airflow is forced through a porous glass medium  (Duran Group, Filter Funnel, porosity 3, inner diameter 56 mm) that is covered by a coarse grid. The bronze grid is made super-hydrophobic (electroless galvanic deposited metal \cite{Larmour:2007} and humid low-surface energy molecular deposition) to avoid imbibition of the hydrophilic porous medium. The large pressure load on the porous medium creates an approximately homogeneous outflow, which is assumed to be hardly affected by the small pressure load of the drop. Consequently, if the airflow $Q$ is large enough, a lubricating layer (air cushion) can emerge and support the complete weight of the drop. There exists a threshold drop size $R$ and gas flow rate $Q$ at which the drops become unstable and start to oscillate, i.e. the instability threshold. The airflow is measured with an Aalborg flow meter (range: 0 - 60 l/min). Since the drop is very mobile in the levitated state, it is necessary to hold it using a needle. This fixates the drop at a constant location on the substrate. The same needle is used to supply and subtract liquid from the drop via a syringe. To study the drop behavior for various flow rates $Q$ and drop sizes $R$, the drop motions are recorded from top view, with a high speed camera at $1000$ fps (Phantom V9). Using a macro lens (Nikon Aspherical Macro, 1:2) with extension tubes, a resolution of 42 $\mu m$/pixel is obtained (see Fig.~\ref{fig:Problemsketch}). Reflective illumination (IDT, LED lightsource) is realised via a 45 degrees tilted beam splitter. 

The aim of this work is to study the instability threshold (appearance of drop oscillations) for levitated drops. To verify reproducibility of the experiment, each measurement is repeated multiple times and by two different procedures. 
In the first method, each measurement starts with a new constant flow rate $Q=Q_t$ and a small drop size $R$. Then the drop volume is slowly increased by pumping liquid into it. The feeding is continued until the drop reaches a floating state ($R < R_{t}$) which finally becomes unstable once the drop size equals the threshold size $R_t$ for flow rate $Q_{t}$. The volume increase of the drop is directly stopped and subsequently, the dynamics of the unstable drop at the threshold value are recorded with the camera. Note that the threshold for levitation and that for the appearance of oscillations are very close to each other.
A second method to determine the instability threshold is measurement of $Q_{c}$, obtained after drops have turned unstable. For a drop starting in the unstable state at $Q=Q_{t}$, the airflow is slowly reduced until a value is reached which results in a stable state: $Q = Q_{c}$. 
This second threshold $Q_c$ turns out to be slightly smaller than $Q_t$.  However, the difference is comparable to the accuracy of the measurements of $Q_c$, so we cannot make any definite statements on whether or not the instability is hysteretic. In what follows we therefore plot the average threshold $Q_m$, obtained upon increasing the drop size and variation of the flow rate. $Q_m$ is determined as: $(Q_t + Q_c)/2$. The error bar indicates the difference between the two measurement procedures.  

After measurement of $Q_c$ the flow rate is further reduced which finally results in a sessile drop state again. A snapshot is made at this zero flow rate (i.e. sessile drop Fig.~\ref{fig:ExampleModes}a) and the drop size $R$ is determined as the maximum radius of the sessile drop in top view. To reduce as much as possible the influence of any possible airflow fluctuations coming from e.g. variations in the substrate or hydrophobic grid fixation, all data points are measured at a fixed position on the substrate. 
To study the influence of viscosity on the drop dynamics, two liquids are used: water ($1~\mathrm{mPa~s}$) and water-glycerine mixture ($60~\mathrm{mPa~s}$). The resulting dynamics are characterized by liquid viscosity $\eta_l$, drop size $R$, flow rate $Q$ and oscillation frequency~$f$.

\section{Experimental results}\label{sec:Experimental results}

\subsection{Low viscosity drops} \label{subsec:lowvisc}

\begin{figure}[htp!]
	\centering 
	\includegraphics[width=0.45 \textwidth]{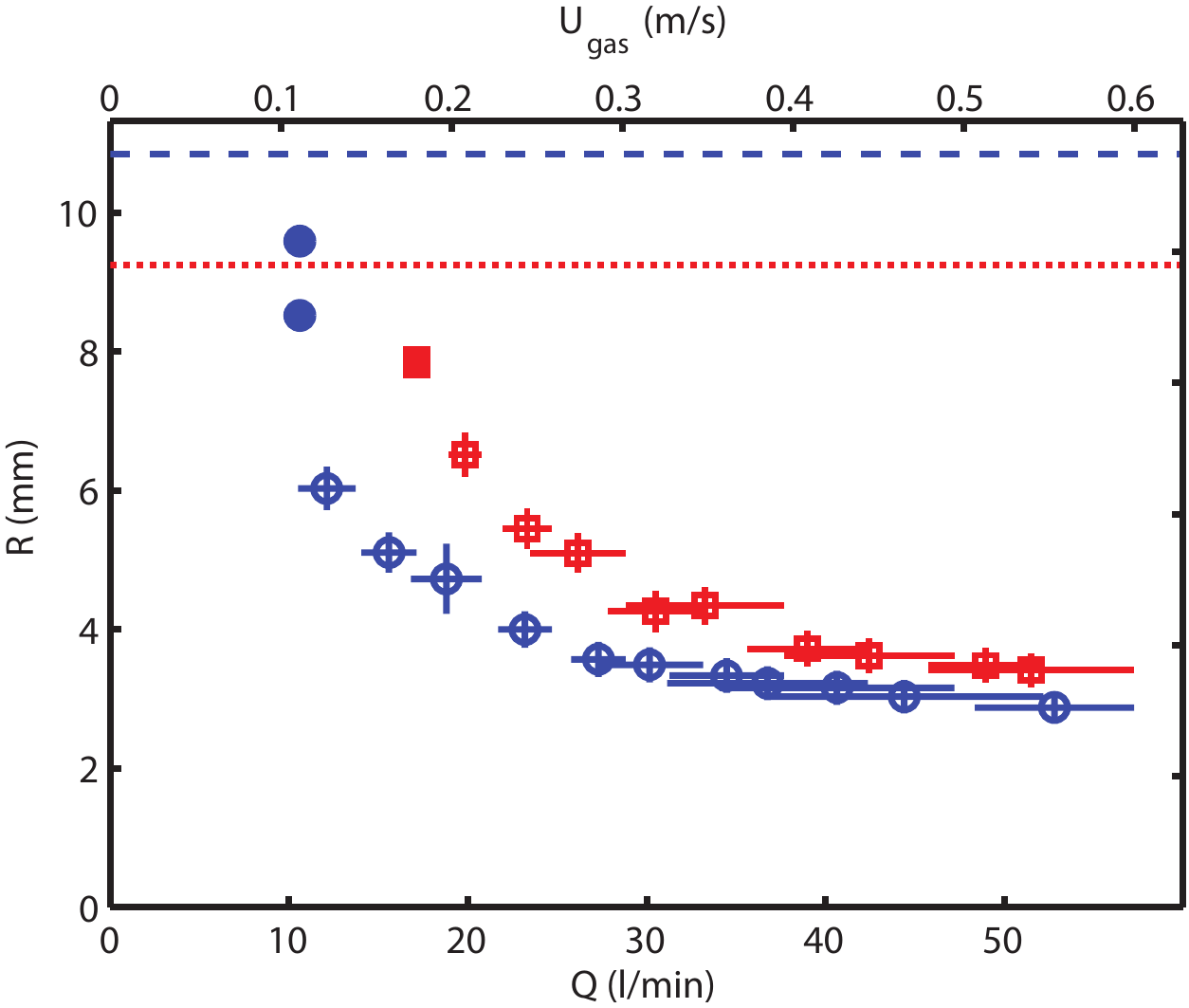}
\caption{Measured instability threshold $Q_m$, for levitated drops. The upper axis gives the gas velocity, estimated by dividing the total flow rate by the area of the porous medium. Data represents all data points for water- and  water/glycerine drops, in circles (\textcolor{blue}{$\circ$} and \textcolor{blue}{$\bullet$}) and squares (\textcolor{red}{$\square$} and \textcolor{red}{$\blacksquare$}), respectively. Since for the smallest flow rate the drop size could not be measured (it detaches from the needle), $R$ is measured in levitated state instead of sessile state. These points are therefore indicated by a solid symbol (\textcolor{blue}{$\bullet$} and \textcolor{red}{$\blacksquare$}). Note that point \textcolor{red}{$\blacksquare$} corresponds to the chimney instability from Fig.~\ref{fig:Breathing}b. The theoretical prediction of the critical radius for chimney instability is indicated by the blue dashed line and red dotted line for the used water and water-glycerine mixture, respectively.}
	   \label{fig:Threshold}
\end{figure}

In this section we study the stability and dynamics of levitated water drops ($\eta_l = 1~\mathrm{mPa~s}$). This is reminiscent to the classical Leidenfrost drops, levitated above a hot substrate \cite{Biance03}. By varying the drop radius $R$ and airflow rate $Q$, the threshold for drop oscillations ($R_t$,$Q_m$) is determined. Results for water are plotted in Fig.~\ref{fig:Threshold}, as circles. The open circles are oscillations without detachment from the needle. In these cases, the size of the drop is measured in sessile state. The solid circles correspond to violent oscillations or a chimney, which can lead to the detachment from the needle. The size is then approximated in the unstable levitated state. Clearly, the threshold drop size $R$ decreases with flow rate. The smallest drops investigated here are stable up to very high flow rate, while the largest drops destabilize even at very small $Q$.  A chimney was for example observed for the smallest flow rate and largest drop size $R\simeq 9.6$mm (top blue solid circle in Fig.~\ref{fig:Threshold}). This point is indeed close to the blue dashed line that indicates the onset of the chimney instability for water drops, as determined for thermal Leidenfrost drops by Biance et al. \cite{Biance03} ($R_c \simeq 4.0 \ell_c$, where $\ell_c$ is the capillary length). Interestingly the chimney instability was predicted to occur even at vanishing flow rate \cite{SnoeijerPRE:2009}. However constraints in the control of extreme small flow rates limited measurements in this range of parameters. 

\begin{figure}[htp!]
	\centering
	\includegraphics[width=0.45 \textwidth]{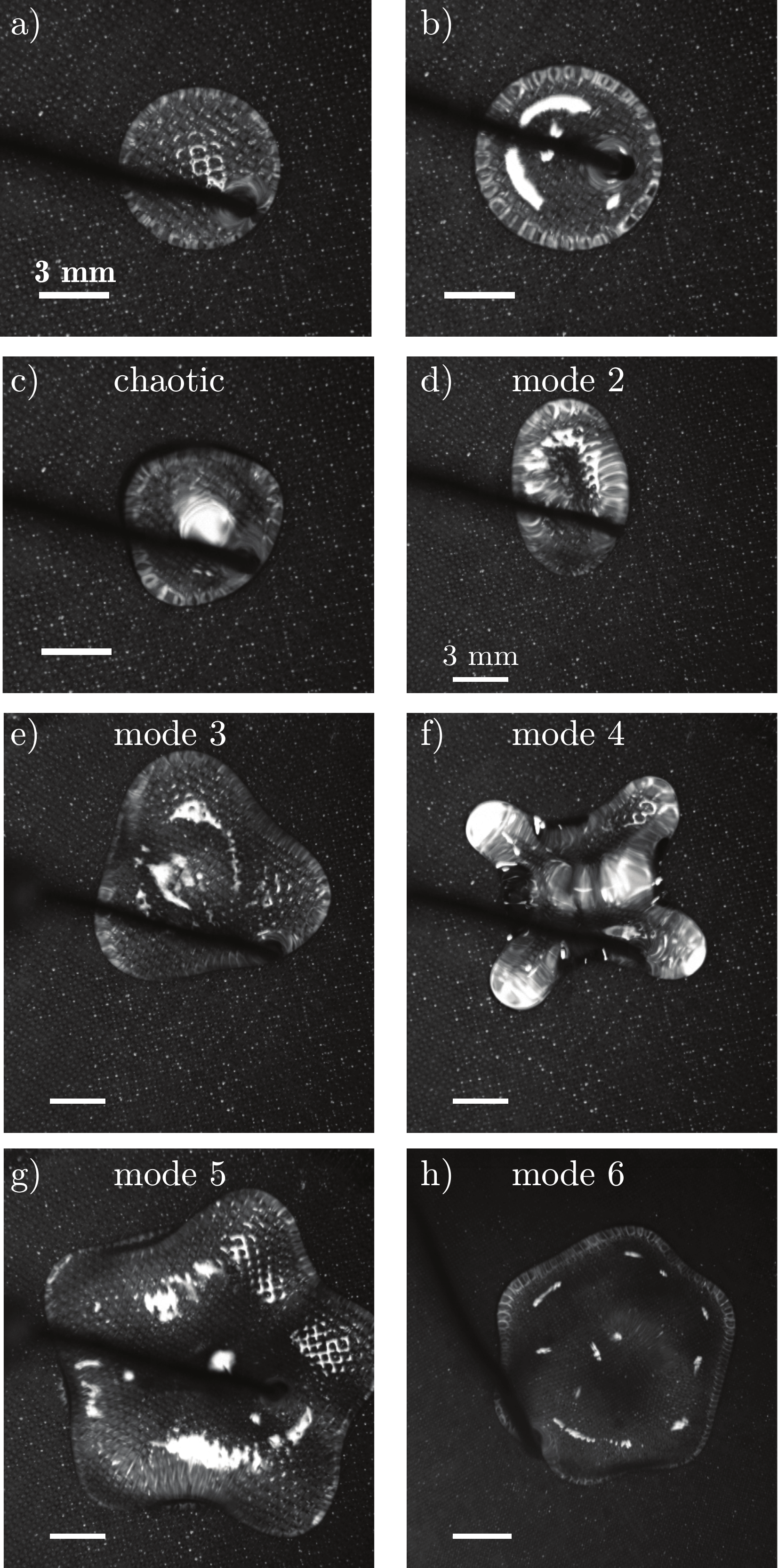}
   \caption{Examples of levitated drop instabilities. All images show water drops, except for (b) which is a water-glycerine drop. (a) Sessile water drop. (b) Levitating water-glycerine drop. (c) Chaotic mode water drop oscillation. (d) Water drop, mode $n=2$ ($R= 4.1$ mm, $f = 13.8$ Hz). (e) Water drop, mode $n=3$ ($R= 6.1$ mm, $f = 14.2$ Hz). (f) Water drop, mode $n=4$ ($R= 5.2$ mm, $f = 17.8$ Hz). (g) Water drop, mode $n=5$ ($R= 8.6$ mm, $f = 14.3$ Hz). (h) Water drop, mode $n=6$ ($R= 6.1$ mm, $f = 30.9$ Hz).}
   \label{fig:ExampleModes}
\end{figure}

For all levitated drops, the oscillating motion is recorded at the threshold flow rate $Q_{t}$.  Typical images obtained in the experiments are shown in Fig.~\ref{fig:ExampleModes}. Fig.~\ref{fig:ExampleModes}a is a sessile water drop, with $Q=0$, while snapshots (Fig.~\ref{fig:ExampleModes}c-h) correspond to oscillating drops at non-zero flow rates. Once the water drops are unstable, the oscillations appear to be rather chaotic, i.e. a combination of modes (Fig.~\ref{fig:ExampleModes}c). However, in few cases also one distinct mode was observed ranging from mode $n=2$ to $n =6$, as is shown in Fig.~\ref{fig:ExampleModes}d-h. 

\begin{figure}[htp!]
\centering
\includegraphics[width = 0.45 \textwidth]{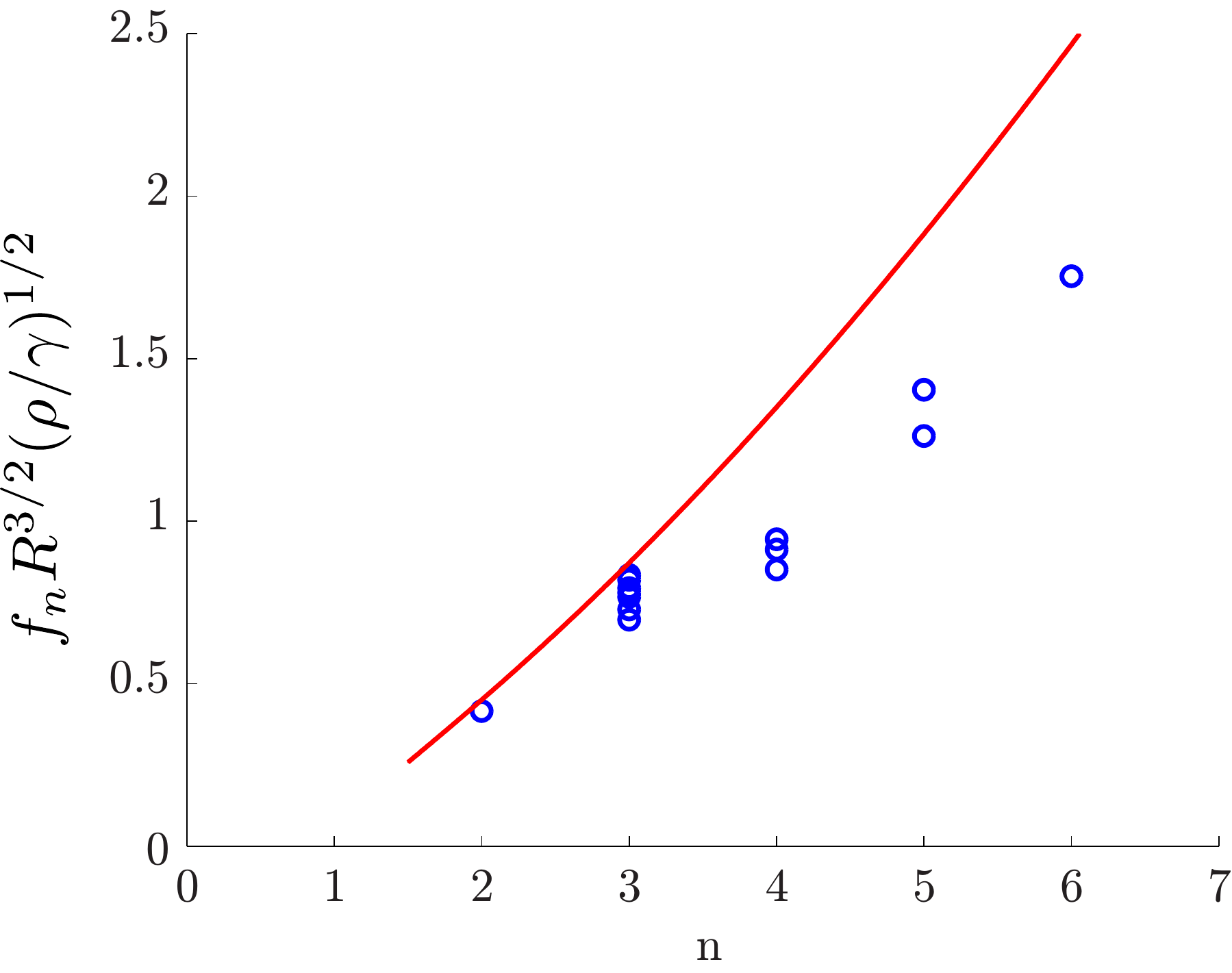}
\caption{The frequency measured for faceted drops as shown in the images of Fig.~\ref{fig:ExampleModes}. Each data point \textcolor{blue}{$\circ$}, corresponds to one water drop measurement. The red solid line is the prediction from the corresponding eigen mode for a puddle, given by eq. (\ref{eq:cylindricalRayleigh}).}
\label{fig:RL}
\end{figure}

In case of these well-defined modes, the oscillation frequency can be determined and compared to the prediction of eq. (\ref{eq:cylindricalRayleigh}). The results are shown in figure~\ref{fig:RL}. For mode $n=3$, frequencies are measured for seven different drop sizes $R=3.2 - 6.1$ mm. Rescaling from eq.~(\ref{eq:cylindricalRayleigh}) indeed collapses the data. Additionally the magnitude and trend are in quite good agreement with the inviscid theory (red solid line) for all modes.

\subsection{High viscosity drops} \label{subsec:highvisc}

\begin{figure*}[htp!]
	\centering
	\includegraphics[width=0.99 \textwidth]{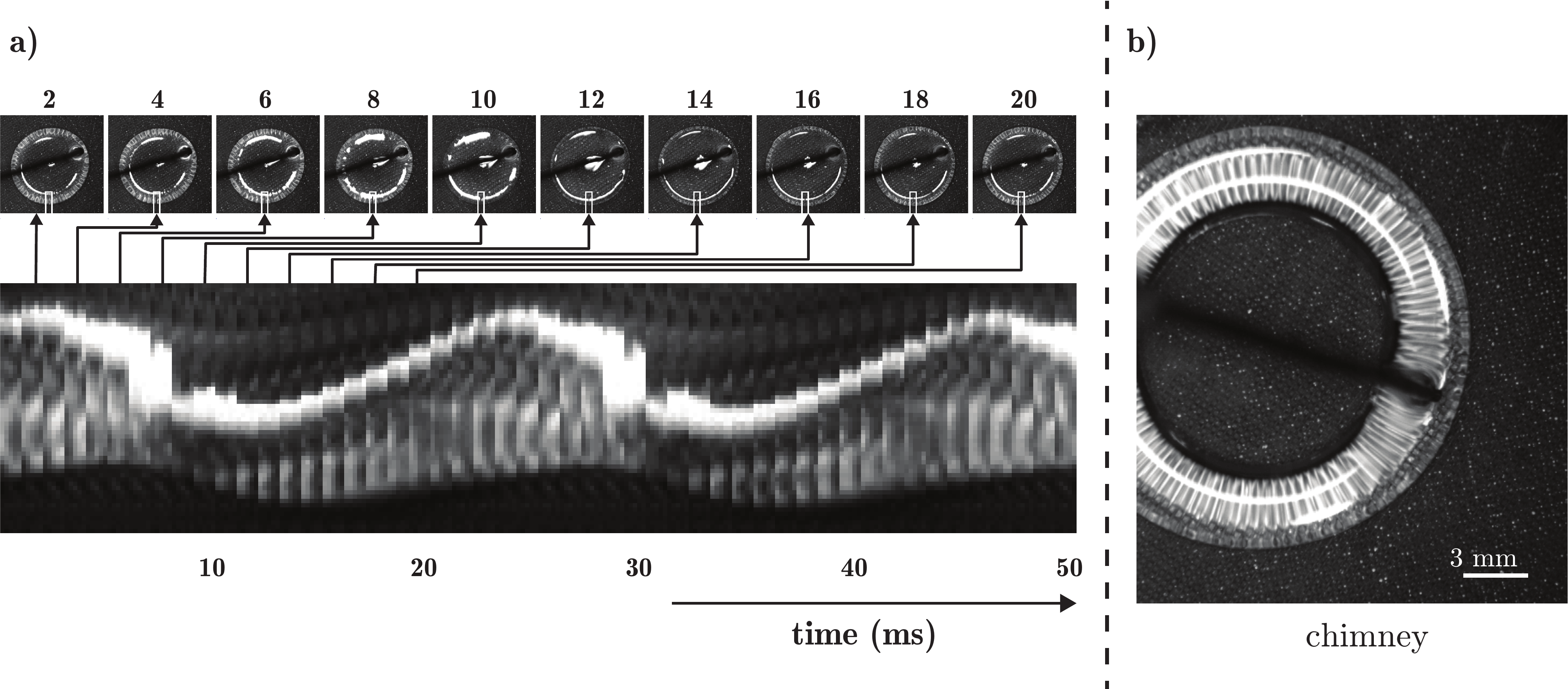}
 	\caption{(a) Top row: an image sequence of the breathing mode oscillation of a large-viscosity drop (water glycerine, $60~\mathrm{mPa~s}$). As the oscillation amplitude is rather small, a space-time diagram is shown as well, which is built from slices similar to the white boxes indicated in the images. (b) For larger drop sizes we observe the formation of a chimney.}
 	  \label{fig:Breathing}
\end{figure*}

The viscosity of the drop is increased to investigate whether damping of the inner drop flow indeed suppresses star oscillations. 
Experiments shown in this section are carried out with liquid drops of water-glycerine mixture ($\eta_l=60~\mathrm{mPa~s}$). Again the drop size $R$ and flow rate $Q$ are varied to determine the instability threshold for drop oscillations. The results are included in Fig.~\ref{fig:Threshold}. The data points for large liquid viscosity are indicated with red squares (\textbf{\textcolor{red}{$\square$}},\textbf{\textcolor{red}{$\blacksquare$}}). For the solid red squared data points, a chimney instability is observed, for which an air bubble pierces through the center of the drop. Such a chimney is shown in Fig.~\ref{fig:Breathing}b. The size of the drop could therefore only be determined from a drop in levitated state.

\begin{figure}[htp!]
	\centering
		\includegraphics[width=0.45 \textwidth]{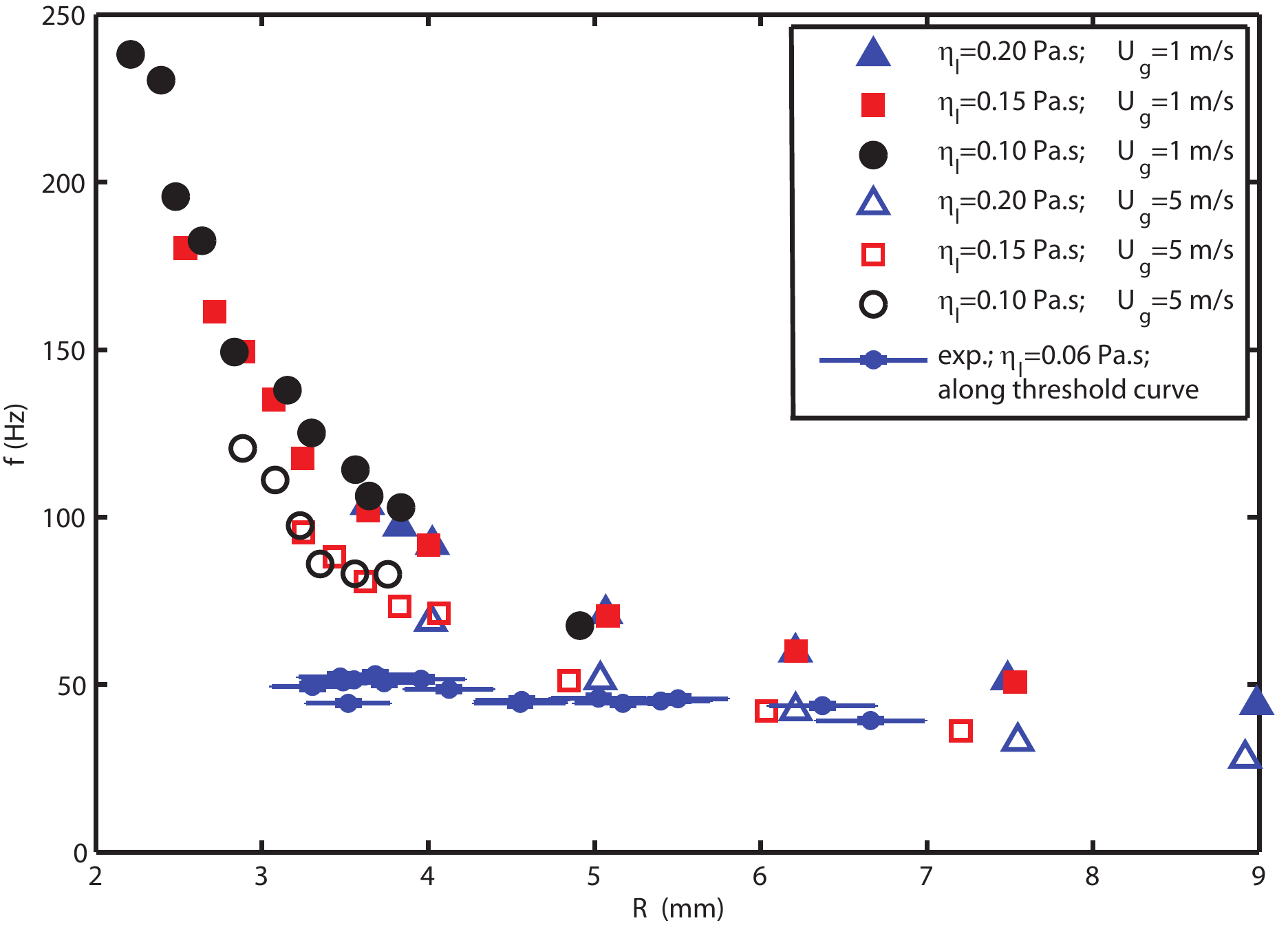}
   \caption{Measured oscillation frequency at threshold for high-viscosity drops (see Fig.~\ref{fig:Threshold}) (blue dots with errorbars), combined with numerical results. For the numerics, the measured oscillation frequency (excitation frequency) as a function of the drop top view radius with airflow velocity 1 and 5 m/s, at three different liquid viscosities is shown. In the numerics, frequency appears to be independent of liquid viscosity, decreases with increasing drop radius, and decreases with increasing airflow velocity.}
   \label{fig:Frequency}
\end{figure}

Comparing the threshold of high viscosity drops with water drops, we observe a clear increase of the threshold. However, the dependence on viscosity is relatively weak, given that the liquid viscosity was increased by a factor of about 60. By contrast, the dynamics are strongly affected by the liquid viscosity. While the oscillations of water drops at threshold is chaotic and non-axisymmetric, the viscous drops only display axisymmetric oscillations: we observe clear `breathing' modes (symbol with error bars in Fig.~\ref{fig:ExampleModes}b), for which the levitated drop remains circular in top view while oscillating. The large viscosity of the liquid drop apparently damps all higher mode oscillations and the formation of star-drops is completely suppressed. A more detailed picture illustrating this dynamics is shown in Fig.~\ref{fig:Breathing}a. Consecutive snapshots (top row) all depict circular drops and a space-time diagram of the drop edge illustrates the radial oscillating motion.
This regular dynamics make it relatively easy to measure the main oscillation frequency for all data along the threshold curve (see Fig.~\ref{fig:Frequency}). Note that in this measurement the frequency therefore is a function of $R(Q_{t})$. Hence, small radius in this figure automatically also means relative large flow rate $Q_{t}$ and vice versa (see Fig.~\ref{fig:Threshold}).

Apart from this large contrast in shape deformations, also the measured oscillation frequencies are different from those measured with low-viscosity water drops. Frequencies for high viscosity drops are considerably higher, by a factor two or more, than the lowest mode (n=2) of the \textit{inviscid} Rayleigh \& Lamb frequency for a drop of the same size, but compare rather well with numerical results for axisymmetric oscillations of an (inviscid) drop on an air cushion (see Sections \ref{sec:Numerical method} and \ref{sec:Numerical results}). One possible interpretation is that the gas flow and the liquid flow act as a coupled dynamic system that oscillates. In case of water this oscillation, acting as a parametric forcing, directly leads to star oscillations which are well described by eq.~\ref{eq:ray}. However, viscosity affects or even suppresses star oscillations in high viscosity drops. As a result one essentially observes the frequency of this axisymmetric oscillation of the coupled system which in contrast to that of the star oscillations only weakly depends on drop size. In summary, due to the suppression of star oscillations viscous drops reveal the underlying axisymmetric oscillation from which the stars originate. It is this axisymmetric oscillation that we will study numerically in the next Sections.

Finally, we again observe chimneys when the drop size becomes too large, $R \approx 8$ mm (see right panel of Fig.~\ref{fig:Breathing}). Since the capillary length for the used water-glycerine mixture is, $\ell_c \sim 2.3$ mm,  the chimney occurs at about $3.5\ell_c$. This is consistent with earlier experiments on water drops~\cite{Biance03} and theory~\cite{SnoeijerPRE:2009} for which the critical radius $R_c \approx 4.0 \ell_c$ ($R_c$ for the water-glycerine mixture is indicated by the red dotted line in Fig.~\ref{fig:Threshold}). 

\section{Numerical method}\label{sec:Numerical method}

We now investigate the dynamics of drops on an air cushion by numerical simulations. Since previous work, where drops were modeled by Stokes flow, did not result into any oscillation \cite{SnoeijerPRE:2009}, inertia inside the drop must be important and we now consider the opposite limit: potential flow. The latter is coupled to a viscous airflow, modeled in the lubrication approximation. The model is similar to that in Ref. \cite{Bouwhuis}, where it was used for simulating drop impact.

\subsection{Parameters \& dimensional analysis}\label{subsec:parameters}

Similar to the experiments, the main parameters that will be varied are the drop volume $V$ and the gas flow, here denoted by the upward gas velocity $U_g$. Other parameters are the gas viscosity $\eta_g$ (lubrication approximation), liquid density $\rho_l$ (potential flow), and the surface tension $\gamma$. These can be combined into three dimensionless numbers. A measure for defining the drop size is the Bond number, $\Bo$, taking into account gravity influence against surface tension influence:

\begin{equation}
~\mathrm{\Bo} = \sqrt{\frac{\rho_lR_0^2g}{\gamma}} = \frac{R_0}{\ell_c},
\label{eq:Bo}
\end{equation}
where $R_0$$=$$\left(\frac{3V}{4\pi}\right)^{\frac{1}{3}}$ is the radius of the unperturbed spherical drop with volume $V$, and $g$ is the acceleration of gravity. $\ell_c$ is the capillary length, as defined in the Introduction. Secondly, we define the capillary number

\begin{equation}
~\mathrm{\Ca}=\frac{\eta_gU_g}{\gamma},
\label{eq:Ca}
\end{equation}
in which $U_g$ is a constant if we assume a uniform upward flow beneath the drop. $\Ca$ measures the influence of gas viscosity against surface tension and can be interpreted as the dimensionless gas velocity.

By setting a balance between the viscous forces of the gas flow and the square root of the inertial forces induced by the drop times the surface tension force, we finally introduce a dimensionless quantity which we will call the Ohnesorge number:

\begin{equation}
~\mathrm{\Oh} = \frac{\eta_g}{\sqrt{\rho_l\gamma \ell_c}}.
\label{eq:Oh}
\end{equation}
Note that this definition of $\Oh$ deviates from the standard definition, since it combines the viscosity of the gas and the density of the liquid.

Then, using $\ell_c$, $\frac{\gamma}{\eta_g}$, and $\frac{\gamma}{\ell_c}$ as the relevant length, velocity and pressure scales, the radial positions $r$, vertical positions $h$, velocities $u$, times $t$, and pressures $P$ are non-dimensionalized as, respectively

\begin{eqnarray}
\widetilde{r}&=&\frac{r}{\ell_c}; \nonumber \\
\widetilde{h}&=&\frac{h}{\ell_c}; \nonumber \\
\widetilde{u}&=&\frac{\eta_g}{\gamma}u; \nonumber \\
\widetilde{t}&=&\frac{\gamma}{\ell_c\eta_g}t; \nonumber \\
\widetilde{P}&=&\frac{\ell_c}{\eta_g}\frac{\eta_g}{\gamma}P = \frac{\ell_c}{\gamma}P. \nonumber
\end{eqnarray}
From now on we will drop the tildes and all variables will be dimensionless, unless stated otherwise.

\subsection{Boundary Integral method coupled to lubricating gas layer}\label{subsec:BI-method and lubrication}

The drop is assumed to consist of an incompressible and irrotational fluid, and can therefore be described by potential flow. The velocity field inside the drop is the gradient of a scalar velocity potential $\phi$. The Laplace equation,

\begin{equation}
\nabla^2 \phi = 0,
\label{eq:Laplace}
\end{equation}
is valid throughout the whole drop including its surface contours. The Boundary Integral method is a way to solve this equation for $\phi$, with the proper boundary conditions \cite{Pozrikidis, Oguz, Bergmann}. For the levitated drop setup, the entire drop surface is a free surface, and the dynamic boundary condition for that surface is the unsteady Bernoulli equation:

\begin{equation}
\frac{1}{~\mathrm{\Oh}^2}\left(\frac{\partial \phi}{\partial t}+\frac{1}{2}\left|\nabla \phi\right|^2\right)=-z-\kappa-P_g,
\label{eq:unsteadyBernoulliequation}
\end{equation}
where $t$ is time, $z$ is the absolute height, and $\kappa$ is the local curvature at a point of the drop surface.  The left-hand side describes the inertial effects of the drop, balanced by gravitational effects, the Young-Laplace pressure, and the influences by the airflow on the right-hand side. $P_g$ is the external pressure which is varying over the lower drop surface after introducing the gas flow. For this, the drop surface has been divided into two parts: the top of the drop where the surrounding pressure is atmospheric; and the bottom of the drop, where we deal with the lubrication pressure induced by the gas flow. The separation point between these two parts is taken at $r=R$, where $R$ is the topview radius, but results are unaffected by the precise location of the division \cite{SnoeijerPRE:2009,Duchemin05}. The gas flow is mainly determined by the viscosity of the gas (Stokes flow). We assume that $R \gg h$. Note that the gas is defined to flow upwards from $z=0$ with uniform gas flow velocity $\Ca$, which will result in a predominantly \itshape radial \upshape gas flow below the drop with velocity $u(r,z)$. For deriving the axisymmetric lubrication approximation, we start with mass conservation of the incompressible gas flow

\begin{equation}
\nabla \cdot \mathbf{u}=0.
\label{eq:continuity_gas}
\end{equation}
Boundary conditions are 

\begin{eqnarray}
u_z|_{z=0}&=&~\mathrm{\Ca}; \nonumber \\
u_z|_{z=h}&=&\dot{h}, \nonumber
\end{eqnarray}
where $\dot{h}$ is the vertical velocity of the drop surface. Furthermore, at the free fluid-air-interface, $z(r)=h(r)$, there is a kinematic boundary condition

\begin{equation}
\frac{\partial h}{\partial t}=u_z |_{z=h}-\frac{\partial h}{\partial r}u_r |_{z=h}, \nonumber
\end{equation}
which is the unsteady part of the problem setting. Integrating the continuity equation (\ref{eq:continuity_gas}) along $z$ (between 0 and $h$), applying Leibniz integral rule, substituting the boundary conditions, defining the average (radial) flow velocity $\overline{u}=\frac{1}{h}\int_0^hu_rdz$, and multiplying the equation with $r$ gives \cite{SnoeijerPRE:2009}

\begin{equation}
\frac{\partial}{\partial r}\left(rh\overline{u}\right)+r\dot{h}=r~\mathrm{\Ca}.
\label{eq:massconservation}
\end{equation}

Applying the Stokes equation for this axisymmetric lubrication flow with zero velocity boundary conditions at $z$=0 and $z$=$h$ gives

\begin{equation}
u=6\overline{u}\left(\frac{z}{h}-\frac{z^2}{h^2}\right) \Rightarrow
\frac{\partial P_g}{\partial r}=-\frac{12\overline{u}}{h^2},
\label{applstokeseq}
\end{equation}
in which $P_g$ is the pressure in the gas layer. Combining (\ref{applstokeseq}) and (\ref{eq:massconservation}), and performing one integration leads to

\begin{equation}
\frac{\partial P_g}{\partial r}=\frac{12}{r h^3}\left(-\frac{\Gamma}{2\pi} + \int_0^r \hat{r}\dot{h}d\hat{r}\right),
\label{eq:lubricationapproximation}
\end{equation}
where

\begin{equation}
\Gamma=2\pi \int_0^r \hat{r}\Ca d\hat{r}=\pi \Ca r^2.
\label{eq:fluxdef}
\end{equation}
is the radius-dependent volume-airflux. The first term on the right-hand-side of (\ref{eq:lubricationapproximation}) is the gas flow term; the second term concerns the motion of the drop interface. $\Gamma$ is radially increasing, since the gas is accumulating beneath the drop.

\subsection{`Artificial' viscous damping}\label{subsec:artificialviscousdamping}

Since viscous effects inside the drop are neglected, all motions (waves, oscillations, vertical translations, ...) will be undamped, as long as we do not apply any form of damping. Indeed, simulations with realistic input parameters (radius and airflow velocity) lead to a quick blow up of surface wave amplitudes or the drop receiving a pressure pulse from below (when $h$ becomes too small at some point). In particular, we were unable to produce any steady solutions without the implementation of damping. We therefore need to introduce a damping term in eq. (\ref{eq:unsteadyBernoulliequation}). We opted to follow a physically motivated way using `Viscous Potential Flow' (VPF) \cite{Joseph03}. Applying VPF to a free surface generally leads to an additional term in the unsteady Bernoulli equation valid on this surface, operating as pure damping term. The additional term is the local normal stress, $2\eta_l \frac{\partial^2 \phi}{\partial n^2}$ \cite{Gordillo}, $\eta_l$ being the liquid viscosity, such that (\ref{eq:unsteadyBernoulliequation}) transforms into:

\begin{equation}
\frac{1}{~\mathrm{\Oh}^2}\left(\frac{\partial \phi}{\partial t}+\frac{1}{2}\left|\nabla \phi\right|^2\right)=-z-\kappa-P_g+2\Lambda \frac{\partial^2 \phi}{\partial n^2},
\label{eq:unsteadyBernoulliequation2}
\end{equation}
where

\begin{equation}
\Lambda = \frac{\eta_l}{\eta_g}.
\label{eq:Lambda}
\end{equation}
We have to make some remarks on this `artificial' damping method. First, it is unclear to what extent the model represents a true viscous drop, since viscosity in general induces vorticity in the flow, which, of course, is absent in the simulation. It turned out that the liquid viscosity required to obtain stable numerical solutions is quite large, about 100 times the viscosity of water. Consequently, we will treat $\Lambda$ as a numerical damping constant, rather than a physical viscous effect of the liquid. Secondly, for too large damping, this method amplifies numerical deviations in the code: the normal stress term contains numerical approximations to derivatives, which are now multiplied by a large factor. Summarizing, both requirements together set a narrow window for our liquid viscosity:

\begin{equation}
0.10~\mathrm{Pa\cdot s} \leq \eta_{l} \leq 0.30~\mathrm{Pa\cdot s}. \nonumber
\end{equation}
Outside this range we were unable to generate reliable and stable numerical results.

\subsection{Numerical details}\label{subsec:Numerical details}

In the numerical process, the Laplace equation is solved every time step, similar to Ref. \cite{Bergmann}. The size of a time step varies over the simulation, and depends on the instantaneous drop dynamics. The time step is small enough to prevent neighboring nodes from crossing each other. For a steady drop, or a falling drop, the time step may be of order 0.001 time units (typically of order $1\cdot10^{-2}~\mathrm{ms}$), while an oscillatory scenario, with strong curvatures and large nodal velocities, could end up with time steps of order $1\cdot10^{-5}~\mathrm{ms}$.

In general, the simulation is initiated by a spherical drop falling from small starting height in the order of 0.10 capillary length. However, close to the chimney instability (see subsection \ref{subsec:Steady shapes and chimneys}), it is necessary to start with a more `gentle' initial shape (i.e. closer to the expected `Leidenfrost' shape for these kind of drop sizes), such that the drop does not get unstable due to the impact of the drop after the free fall.

The drop contour is characterized by $r$ and $z$ for $r>0$. For the initial spherical drop (in the first time steps of the simulation), this surface line consists of about 60 nodes, depending on the size of the drop (a smaller drop results in a smaller number of nodes). The number of nodes will vary during the simulation, set by the (maximum) local curvatures on the line and the closeness to the symmetry-axis $r$=0; the largest node density is set around the bottom and top center of the drop. It has been checked that further increasing the number of nodes does not change the results significantly.

\section{Numerical results}\label{sec:Numerical results}

To easily compare with experiments, the figures in this section are in SI units.

\subsection{Steady shapes \& chimneys}\label{subsec:Steady shapes and chimneys}

\begin{figure}[htp!]
\centering
\includegraphics[width=8.0cm]{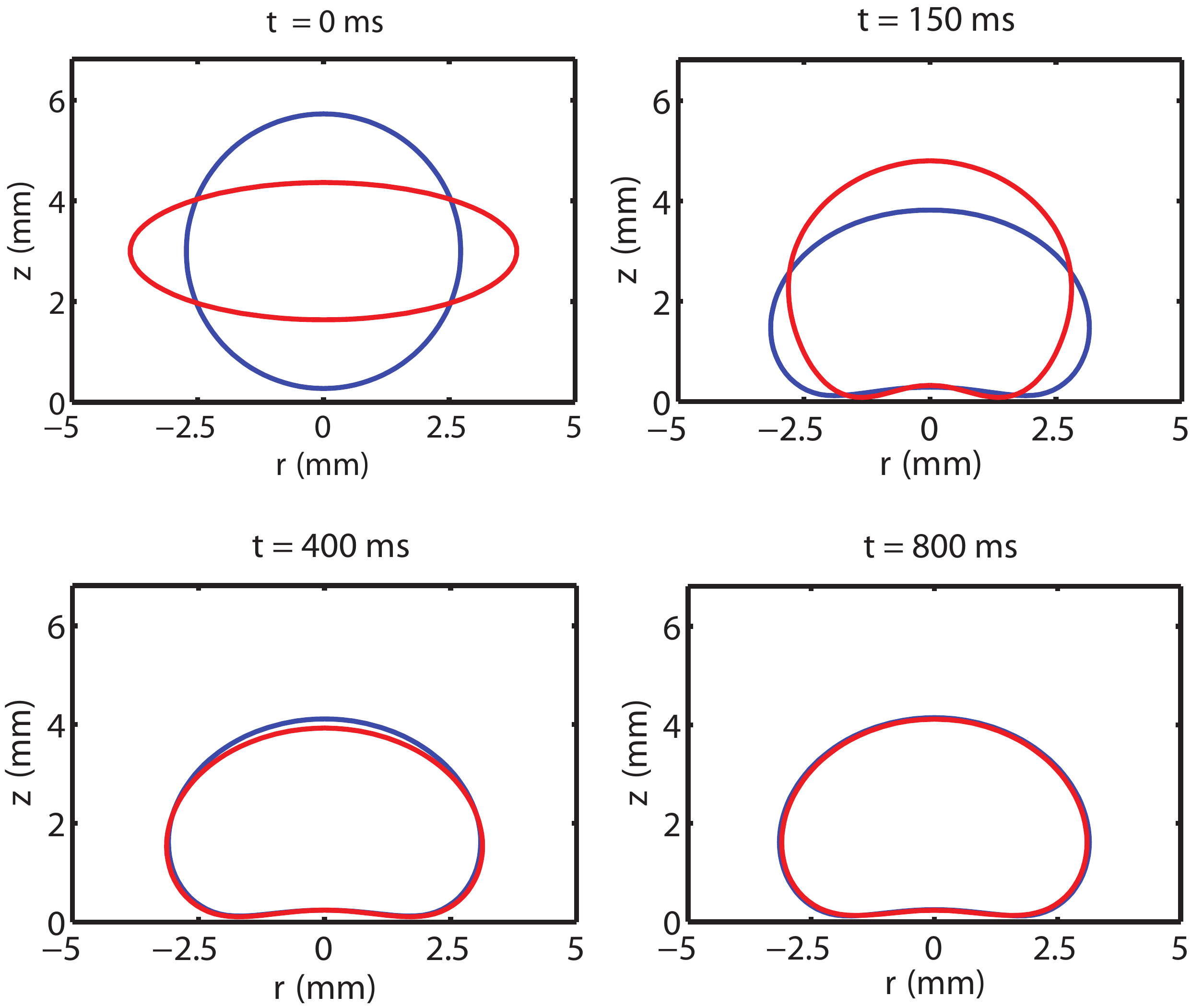}
\caption{Two different initial drop shapes (one spherical, one elliptical) of equal volume, converging to the same steady end shape. $\Bo$=1, $\Ca$=2.5$\cdot10^{-4}$, $\Lambda$=11$\cdot10^3$. See Fig. \ref{fig:pressureprof} for the corresponding pressure profile.}
\label{fig:Equilibriumshape}
\end{figure}

\begin{figure}[htp!]
\centering
\includegraphics[width=8.0cm]{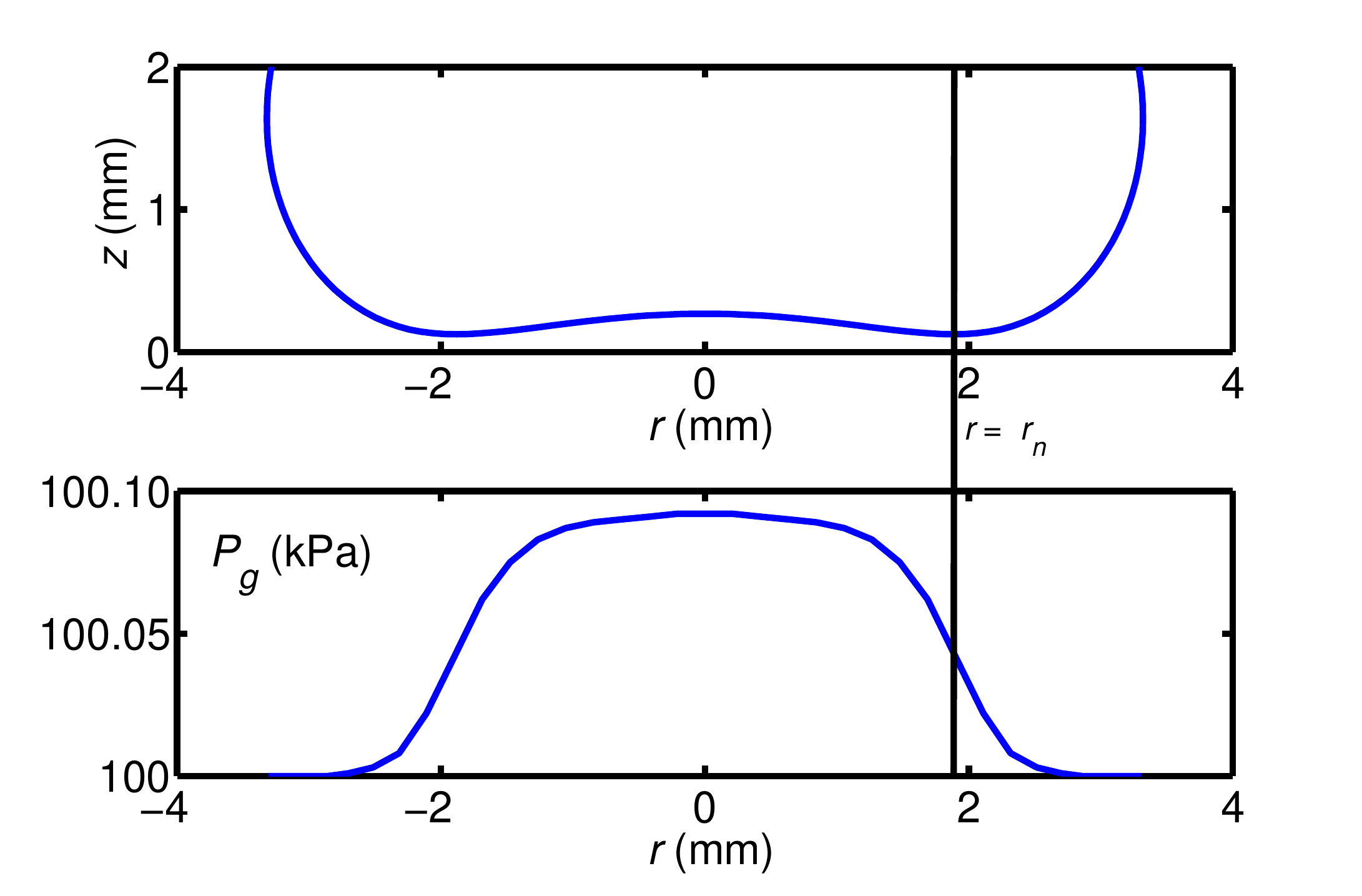}
\caption{Pressure profile ($P_g$) at the bottom of a steady levitated drop. The largest pressure gradient is typically at the neck, $r=r_n$, such that the profile resembles a plateau. $\Bo$=1, $\Ca$=2.5$\cdot10^{-4}$, $\Lambda$=11$\cdot10^3$}
\label{fig:pressureprof}
\end{figure}

\begin{figure}[htp!]
\centering
\includegraphics[width=8.0cm, height=5.0cm]{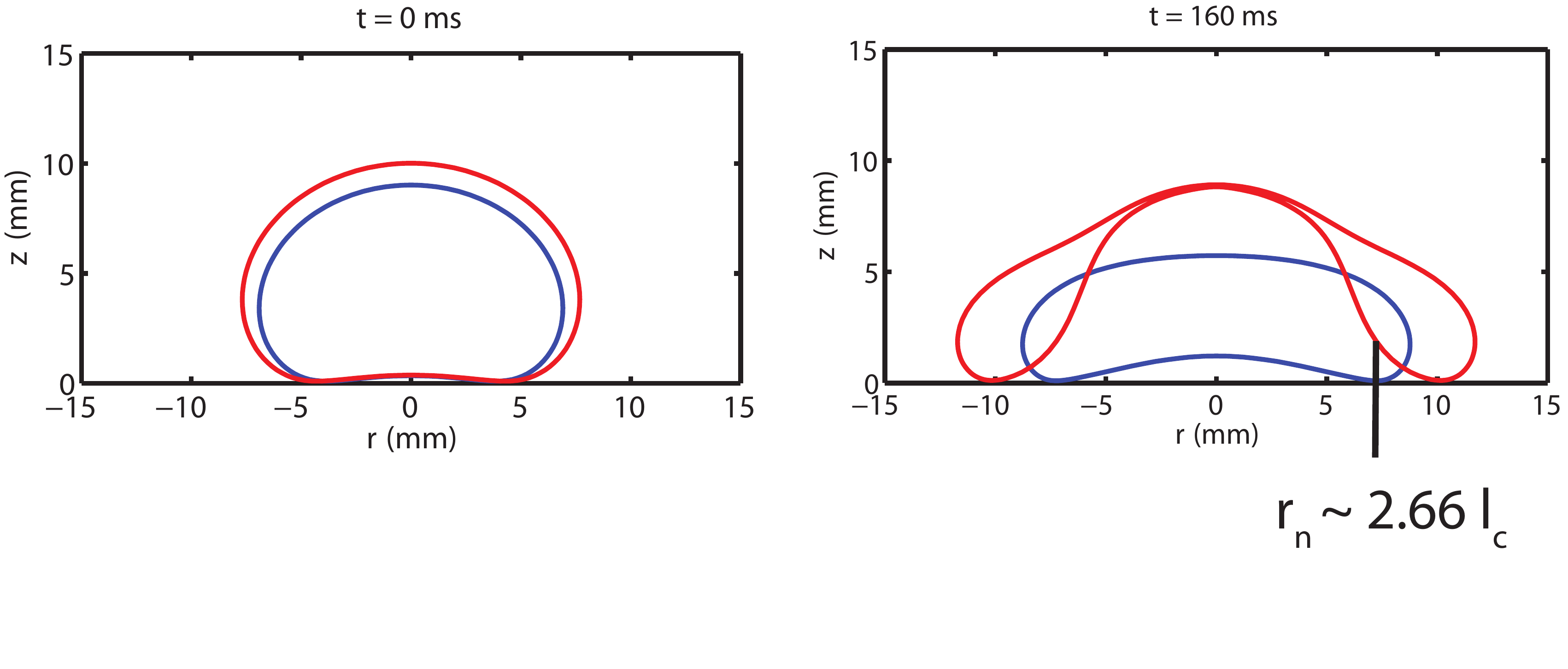}
\caption{Chimney instability. Shown is the evolution of two drops, with \emph{almost} equal volumes. The largest drop apparently has a radius just above the chimney threshold, which appears to be about 2.7$\ell_c$, or 7.3 mm, while the smallest has a radius just below. The large drop does not remain stable, due to the gas pocket breaking through; the small drop keeps its steady shape. $\Bo$=2.25 and 2.5, $\Ca$=2.5$\cdot10^{-5}$, $\Lambda$=11$\cdot10^3$.}
\label{fig:Chimney}
\end{figure}

The numerical scheme described above can indeed lead to steady levitated drops, chimneys, or oscillatory states, depending on the model parameters. Here we first focus on steady shapes, an example of which is shown in Fig. \ref{fig:Equilibriumshape}. For two different initial conditions (top left panel), the drop relaxes to the same final shape (bottom right panel). In all cases, the drop shape depends only on $\Bo$ and $\Ca$, and is independent of $\Oh$ and $\Lambda$.

The pressure profile at the bottom of the drop has a similar shape for every drop size and airflow velocity, from the moment the steady shape has been reached. An example is shown in Fig. \ref{fig:pressureprof}. The largest pressure is at $r=0$, and it decreases to atmospheric pressure for $r\rightarrow R$. The pressure gradient is largest at the neck radius, $r=r_n$, such that the pressure profile resembles a plateau. The minimal gap height in this example is of the order of $100~\mathrm{\mu m}$.

Fig. \ref{fig:Chimney} shows an example of a chimney instability. The respective volumes of the red and blue curves differ by a small amount. Yet, the bigger drop develops a chimney instability, while the smaller one exhibits a steady state. The limit of drop size for the chimney instability agrees with expectations from Ref. \cite{SnoeijerPRE:2009}. We deduce from Fig. \ref{fig:Chimney} a threshold neck radius of about 2.7$\ell_c$ for a gas flow velocity of $0.1~\mathrm{m/s}$. The dimensionless airflux $\chi$ which is introduced in Ref. \cite{SnoeijerPRE:2009} is in our case $\chi=\frac{6\Gamma(r_n)}{\pi r_n}=\frac{6\pi \cdot \Ca \cdot r_n^2}{\pi r_n} \sim \frac{6\pi0.1(2.7\ell_c)^2}{\pi(2.7\ell_c)}=4.42$ $\cdot$ $10^{-3}$. Extrapolation in Fig. 12 of Ref. \cite{SnoeijerPRE:2009} shows that this $2.7\ell_c$ agrees with the theoretical prediction coming from the lubrication approximation. The threshold for chimneys is at smaller drop size than the experimentally observed threshold (Fig. \ref{fig:Threshold}), which can be explained by the smaller incoming airflow velocity in the experiments, compared to numerics. According to Ref. \cite{SnoeijerPRE:2009}, for increasing $\chi$, the threshold for chimneys is at smaller drop size, and $\chi$ in the numerics is indeed large with respect to $\chi$ in the experiments.

\subsection{Drop oscillations}\label{subsec:Drop oscillations}

\subsubsection{Observations}\label{subsubsec:Observations}

\begin{figure*}[htp!]
\centering
\includegraphics[width=14.0cm, height=7.0cm]{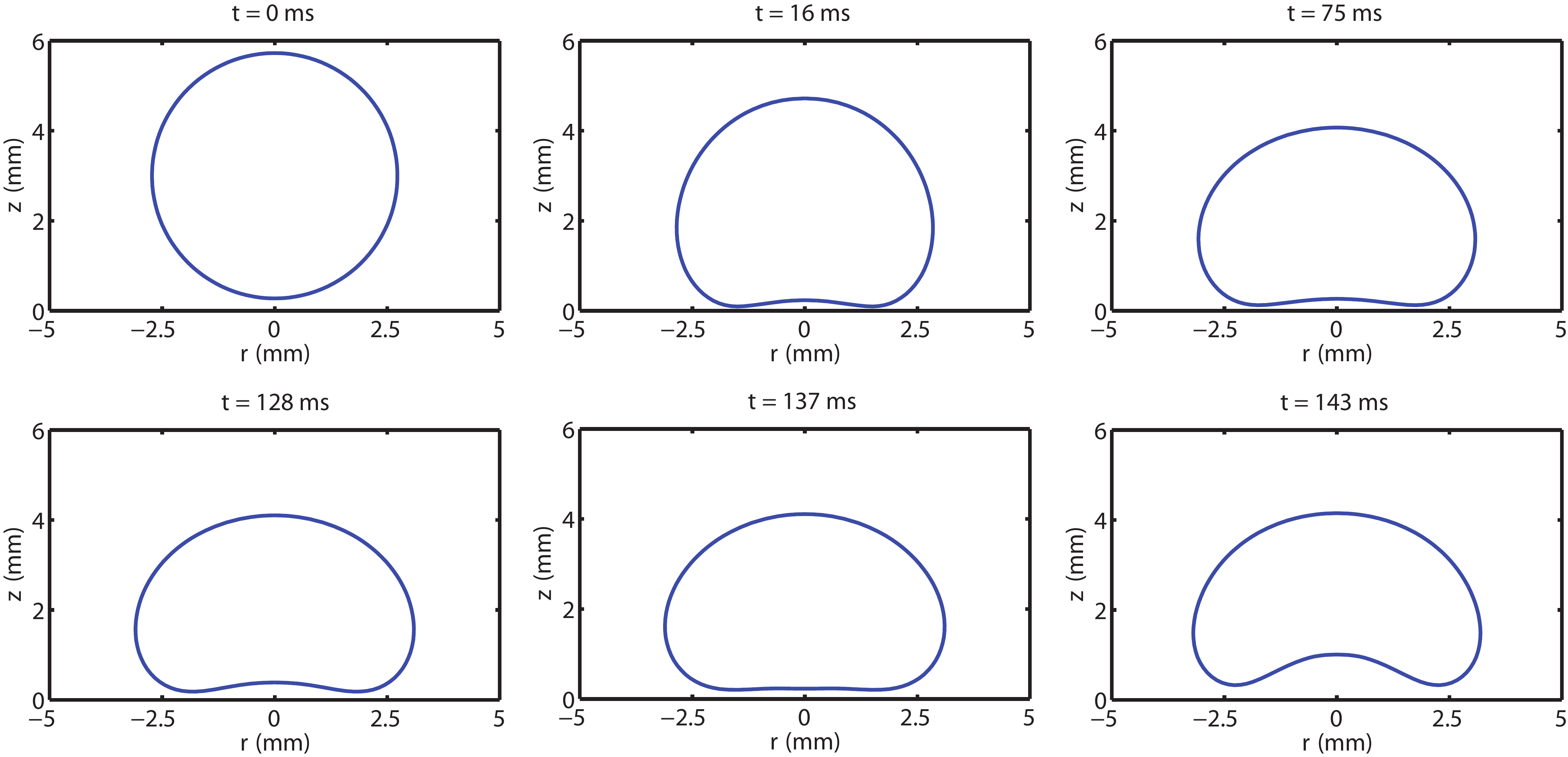}
\caption{Time sequence from the evolution of the oscillatory instability of a levitated drop. The simulation is initiated by a spherical drop, released from small height (0.27 mm) (top-left). The top panel row shows the process from the spherical drop shape to an intermediate steady shape. The bottom panel row shows the oscillatory behavior of the drop at a later point in time. $\Bo$=1, $\Ca$=2.5$\cdot10^{-4}$, $\Lambda$=8.2$\cdot10^3$.}
\label{fig:Timesequence}
\end{figure*}

\begin{figure}[htp!]
\centering
\includegraphics[width=8.0cm, height=8.0cm]{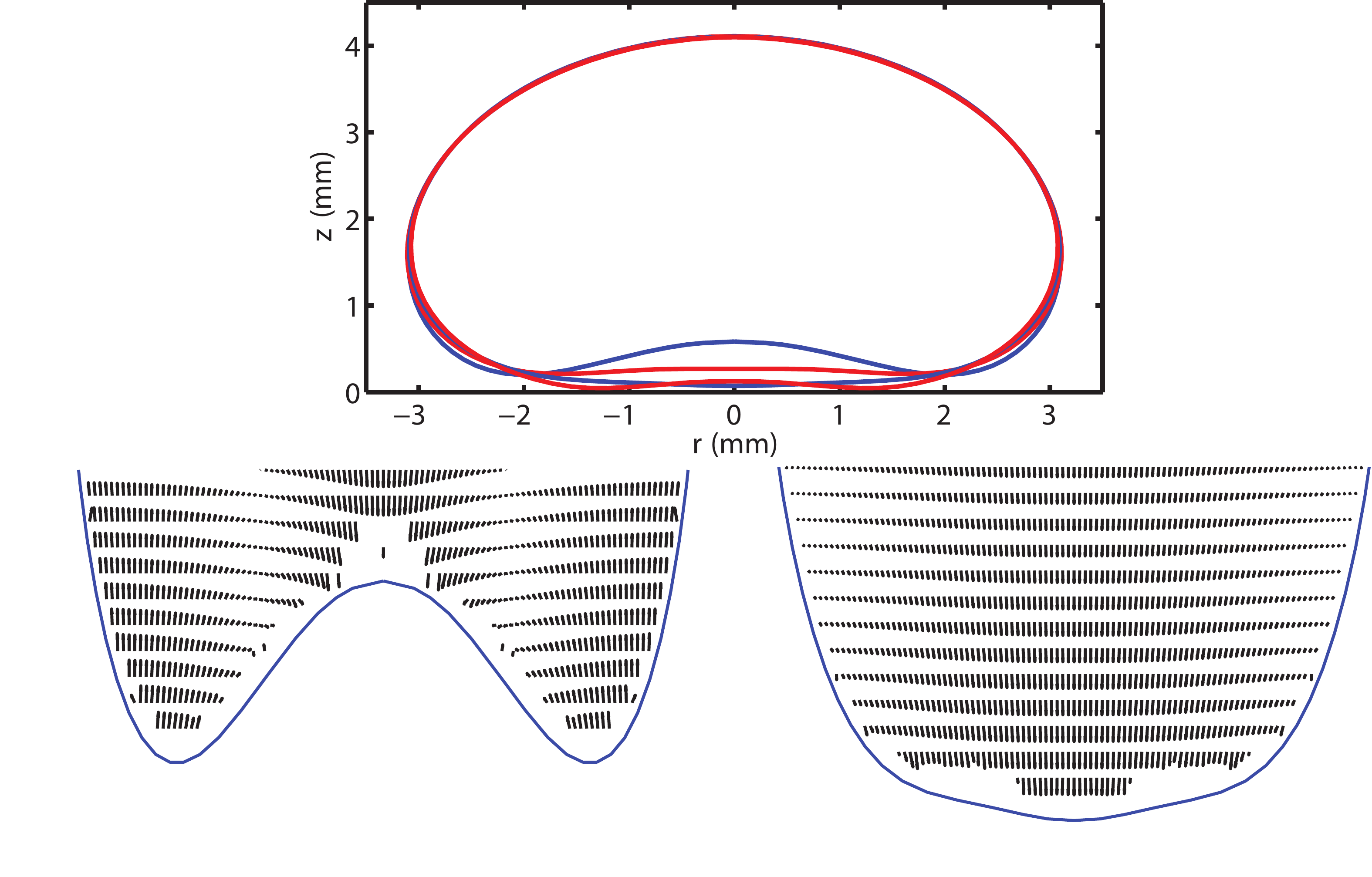}
\caption{Drop contours during the final stages of the simulation in an unstable scenario (see Fig. \ref{fig:Timesequence}). Blue contours are the two extremes, red lines are intermediate. The bottom two plots show the velocity profile inside the drop for the extremes. Note that the liquid velocities, and therefore the oscillations as well, are mainly in the vertical direction. $\Bo$=1, $\Ca$=2.5$\cdot10^{-4}$, $\Lambda$=8.2$\cdot10^3$.}
\label{fig:contours}
\end{figure}

\begin{figure*}[htp!]
\centering
\includegraphics[width=17.0cm, height=4.0cm]{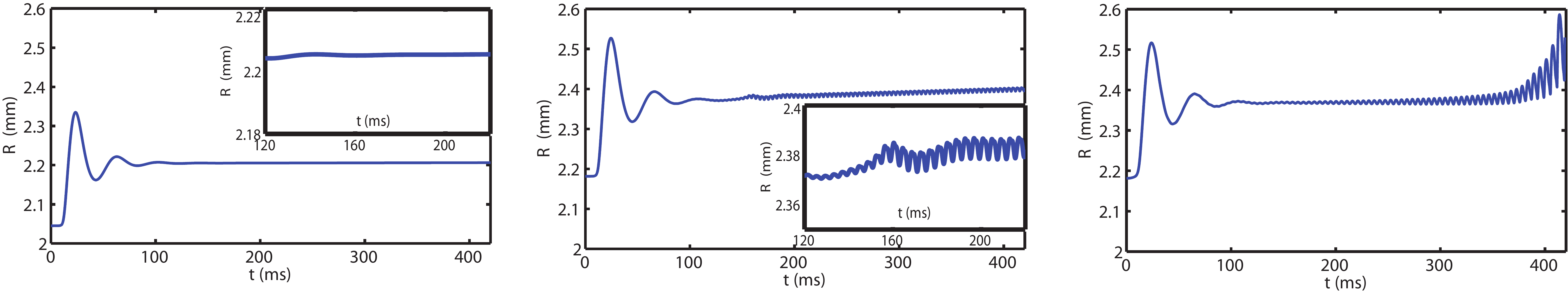}
\caption{Top view radius as a function of time for (a) a stable drop, (b) a case around the transition, and (c) an unstable drop. In the first part of each plot (up to about $100~\mathrm{ms}$) the initial, spherical shape of the drop stabilizes towards the `Leidenfrost' state. After this stabilization the oscillations become visible which typically have a much larger frequency (see insets). (a) $\Bo$=0.75, $\Ca$=2.5$\cdot10^{-5}$; (b) $\Bo$=0.80, $\Ca$=5$\cdot 10^{-5}$; (c) $\Bo$=0.80, $\Ca$=5$\cdot 10^{-4}$. $\Lambda$=5.5$\cdot 10^3$ in all three cases.}
\label{fig:Oscillations}
\end{figure*}

The second scenario of interest we studied is drop instability leading to oscillations. An example is shown in Fig. \ref{fig:Timesequence}, showing the drop contours during the evolution of the oscillations for an unstable scenario. The first three panels (top row) show the process of the drop converging towards the `Leidenfrost' shape. It takes about 75 ms for the drop to adopt a nearly steady shape (top-right), but in the next phase surface oscillations with increasing magnitude are visible (bottom sequence). The drop oscillates in both radial and vertical direction. The two states between which the drop `bounces' are clearly visualized in the last two frames of Fig. \ref{fig:Timesequence}, and in Fig. \ref{fig:contours}, supplemented with velocity profiles. The velocity profiles show that the liquid velocity, and therefore the oscillations and momentary liquid flows are mainly in the vertical direction. Air is released from the gas-pocket at the bottom of the drop around one of the extremes and is gathered again towards the other: the system `breathes'.

Similarly to experiments, there exists a drop size threshold and a gas flux threshold above which the surface oscillations appear. In Fig. \ref{fig:Oscillations}a, no drop oscillations are visible. In Fig.~\ref{fig:Oscillations} we plot the time dynamics $R(t)$ for different parameters. In Fig.~\ref{fig:Oscillations}b, the oscillation amplitude visibly saturates at some small level. The threshold for oscillations is determined for the smallest asymptotically detectable oscillation. In Fig.~\ref{fig:Oscillations}c, the oscillation amplitude starts to grow after some time and the drop does not reach any asymptotic state, which is clearly an unstable situation. This explosive scenario is observed at some distance beyond the oscillatory threshold. The growth rate of the instability depends on the gas flux and the drop size, but especially on the damping coefficient $\Lambda$.

\subsubsection{Stability diagram}\label{subsubsec:Stability diagram}

\begin{figure}[htp!]
\centering
\includegraphics[width=8.0cm]{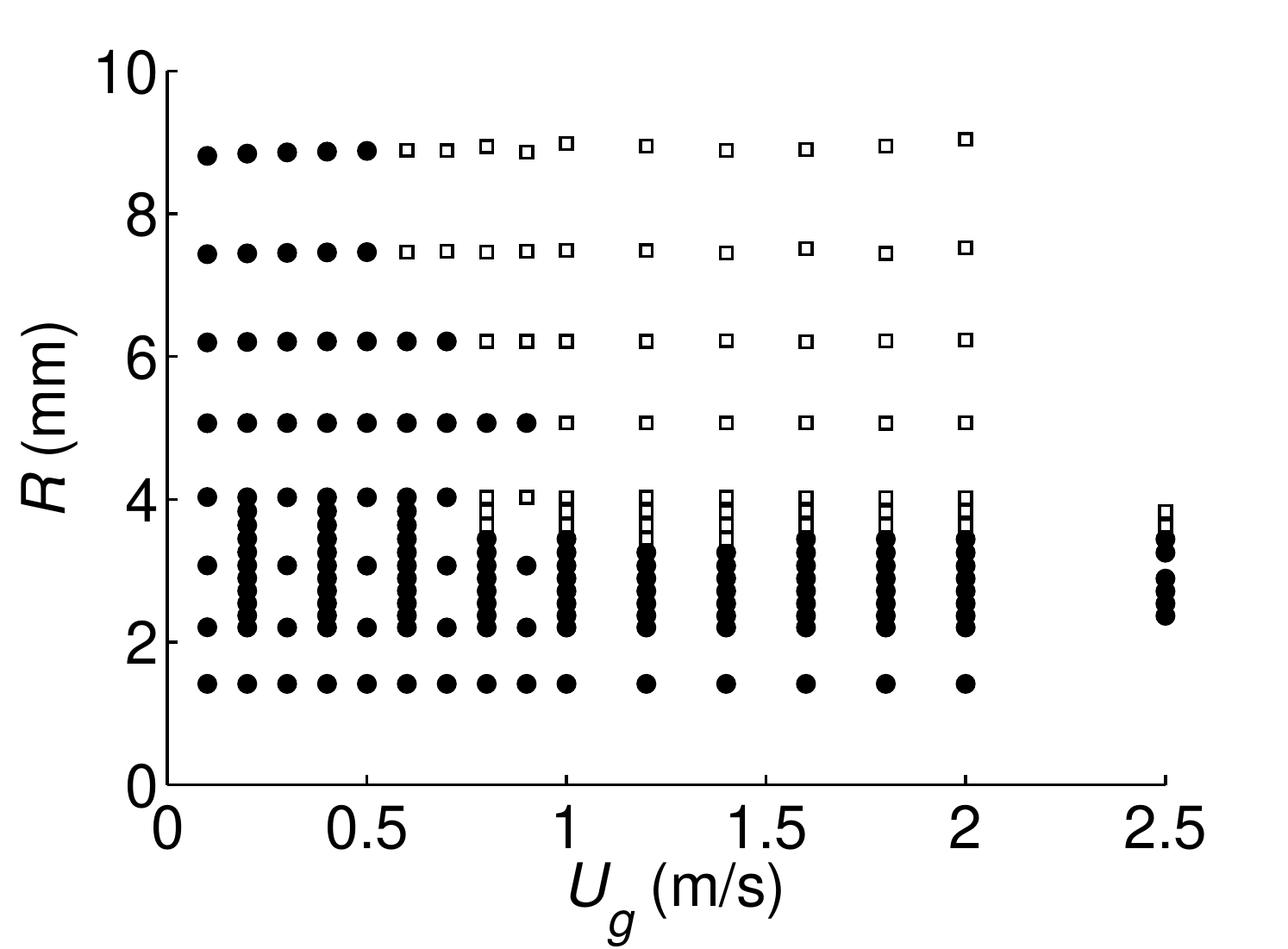}
\caption{Top view radius $R$ versus airflow velocity $U_g$ stability diagram for $\Lambda$=11$\cdot 10^3$. Black dots point out stable configurations: the drop has no tendency to oscillate; white squares indicate oscillating drops.}
\label{fig:Stabilitydiagram_mu=0dot20}
\end{figure}

We investigated the threshold for obtaining surface oscillations by varying the drop size and the gas flow velocity for $\eta_l$=$0.20~\mathrm{Pa \cdot s}$, resulting in the stability diagram shown in Fig. \ref{fig:Stabilitydiagram_mu=0dot20}. We observe a decreasing transition line, similar to the experimental results in Fig. \ref{fig:Threshold} with larger drops becoming unstable at smaller airflow velocity. An important observation is that the threshold is at much larger values (approximately a factor of 10 larger) for the ascending airflow velocity (factor of about 10), compared to the experiments (see Fig. \ref{fig:Threshold}). The relative shape of the transition line is similar in all numerical stability diagrams obtained for different $\eta_l$ and $\rho_l$, but for decreasing damping factor and/or increasing liquid density, the line moves in both the left and the downward direction. In experiments, the influence of the liquid viscosity on the threshold of the instability turned out to be very small. Obviously, our artificial implementation of damping is a plausible reason for the discrepancy between experiments and numerics concerning the threshold.

\subsubsection{Frequency analysis}\label{subsubsec:Frequency analysis}


In Fig. \ref{fig:Frequency}, we show the measured drop oscillation frequencies from the simulations against the drop radius, for different $\eta_l$ and $U_g$, and compare them to the experimental values for a water-glycerine drop. The oscillation frequencies decrease with increasing drop size, and decrease slightly with increasing gas flow velocity. The observed frequencies appear to be independent of the damping factor. 

The frequencies extracted from numerics are compared to those measured experimentally on axisymmetric oscillations for highly viscous drops: The agreement is good for the large radii ($R$ from 5 to 7 mm), but there are some discrepancies for smaller drop radius. To understand this overestimation from numerics, it should be pointed out that the magnitude of oscillations can be much larger in experiments than in numerics. Non-linear effects at finite amplitude generally lead to a decrease of the response frequency of drops \cite{Smith10}, which is especially prevalent for small drops.

\section{Discussion}\label{sec:Discussion}

In this paper we investigated the dynamics of drops levitated by a gas cushion with constant and uniform influx. Various dynamics are observed, both in experiments and numerics: Drops either exhibit stable shapes, oscillate, or, undergo a `chimney' instability in which the gas pocket breaks through the center of the drop.

Our experimental results show that for both high-viscosity and low-viscosity drops, the threshold flow rate for oscillatory instability continuously increases when decreasing the drop size. At very low $Q$, we do not reach the oscillatory state, since there is a maximum drop size beyond which the chimney instability sets in, as predicted by Snoeijer et al.~\cite{SnoeijerPRE:2009}. The trends are very similar for both viscosities, but the threshold is slightly higher at high viscosity. This dependence on viscosity is relatively weak in our experiments; whereas the viscosity was increased by a factor 60, the threshold flow rate only increased by less than 50\%. By contrast, the drop dynamics \emph{are} strongly influenced by viscosity. Non-axisymmetric modes and chaotic oscillations could be observed near the threshold in oscillating water drops, while in the high viscosity case, \textit{only} the `breathing' mode is observed. From this observation we infer that axisymmetric modes rather than the breaking of the azimuthal symmetry constitute the origin of the spontaneous appearance of oscillations.

All these features have been reproduced numerically, by coupling inviscid Boundary Integral code for the drop to a viscous lubrication model for the gas flow. Because potential flow without any damping was unstable in the interesting time range for the evolution of drop oscillations, an artificial damping needed to be introduced, which enabled the observation of both stable drop shapes and oscillations. The idea of a coupling between potential flow liquid and Stokes gas flow proved to be very useful to study the equilibrium shapes of Leidenfrost drops and deforming dynamics of these drops, or (the dimple formation of) impacting drops at room temperature \cite{Bouwhuis} and impacting evaporating drops. Interestingly, for the impacting drop simulations, no damping needed to be involved (because the time range in which we are interested was much shorter).

In the numerical simulations of Leidenfrost drops it is observed that, within a certain range of the parameter space, initially stable (steady) drop shapes gradually start to oscillate. Frequencies of the oscillations are in reasonable agreement with experimental results, especially for large drops. The most important difference between our numerics and the experiments is that the threshold strongly depends on the amount of damping, and that the threshold velocity lies an order of magnitude away from the experimental one. Therefore, a more realistic way of damping needs to be implemented to investigate the position of the threshold.

In both experiments and simulations, the air is injected from below. This is different from Leidenfrost drops, which float on their own vapor, but their dynamics are very similar. Hence, it is verified that the phenomenon of star oscillations does not require any thermal driving, contrarily to previous suggestions \cite{Japonais94}. This confirms the preliminary experimental observation~\cite{Brunet11} that the origin of drop oscillations are purely governed by fluid dynamics. The picture that emerges is that the oscillations appear due to an instability of the coupled system of the lubricating gas flow and the deformable drop. In the experiments, once the oscillations appear, `stars' naturally develop as a parametric instability for low-viscosity drops, in a way similar to water drops placed on an oscillating plate~\cite{Japonais96}. At higher viscosity, the star formation is suppressed by viscous damping and only axisymmetric modes appear. This is similar for the onset of Faraday waves, induced by periodic forcing of a horizontal free-surface \cite{Kumar_Tuckerman}. Indeed, a large viscosity suppresses  the appearance of the parametric instability that leads to Faraday waves. Therefore, this confirms that faceted star shapes are a result of parametric excitation that can only appear at sufficiently small damping (i.e. liquid viscosity).

Though the exact mechanism that leads to oscillations remains to be explained, our study unveiled interesting clues to understand the phenomenon and could dismiss other mechanisms. Interestingly, the Reynolds number for the high viscosity drops in experiments is relatively small $\Re_l \sim \widetilde{U}_l R \rho_l/ \eta_l \sim 0.1 R f R \rho_l / \eta_l \approx 1$ (where we estimate the oscillation amplitude as 10$\%$ of $R$) and still spontaneous oscillations are observed above a threshold radius and gas flow rate. Previous numerical simulations based on Stokes flow for both the drop and the gas displayed no oscillations \cite{SnoeijerPRE:2009}. This raises the question of whether oscillations indeed cease to exist when further reducing the Reynolds number, i.e. by increasing the liquid viscosity. It will be a challenge to investigate this regime experimentally due to practical difficulties of working with such a highly viscous liquid. Other valuable information could also be provided by flow visualization inside the drop and the gas, since the results suggest a crucial coupling between the drop flow and the gas flow. The latter method does not only apply to the experiments, but also to the numerics.

\section*{Acknowledgements}

This work was funded by VIDI Grant. No. 11304 and is part of the research program `Contact Line Control during Wetting and Dewetting' (CLC) of the `Stichting voor Fundamenteel Onderzoek der Materie (FOM)', which is financially supported by the `Nederlandse Organisatie voor Wetenschappelijk Onderzoek (NWO).' The 1 month stay of K.G. Winkels at MSC laboratory was partly funded by ANR Freeflow.

\end{document}